\documentclass[aps,prl,twocolumn,epsfig,amsmath,showpacs,superscriptaddress]{revtex4-1}

\usepackage{epsfig}
\usepackage{dcolumn}
\usepackage{bm}
\usepackage{color}

\usepackage{hyperref}
\hypersetup{
 pdfnewwindow=true, colorlinks=true,
 linkcolor=blue, anchorcolor=blue,
 citecolor=blue, filecolor=blue,
 menucolor=blue, urlcolor=blue}

\begin{document}
\title{Inhibiting Klein tunneling in graphene p--n junction without an external magnetic field}

\author{Hyungju Oh}
\email[Email:\ ]{xtom97@civet.berkeley.edu}
\affiliation{Department of Physics, University of California, Berkeley, California 94720, USA} 
\affiliation{Materials Sciences Division, Lawrence Berkeley National Laboratory, 
Berkeley, California 94720, USA}
\author{Sinisa Coh}
\affiliation{Department of Physics, University of California, Berkeley, California 94720, USA} 
\affiliation{Materials Sciences Division, Lawrence Berkeley National Laboratory, 
Berkeley, California 94720, USA}
\author{Young-Woo Son}
\affiliation{Department of Physics, University of California, Berkeley, California 94720, USA} 
\affiliation{Materials Sciences Division, Lawrence Berkeley National Laboratory, 
Berkeley, California 94720, USA}
\affiliation{Korea Institute for Advanced Study, Seoul 130-722, Korea}
\author{Marvin L. Cohen}
\affiliation{Department of Physics, University of California, Berkeley, California 94720, USA} 
\affiliation{Materials Sciences Division, Lawrence Berkeley National Laboratory, 
Berkeley, California 94720, USA}

\date{\today}

\begin{abstract}
We study by first-principles calculations a densely packed island of
organic molecules (F$_4$TCNQ) adsorbed on graphene.  We find that with
electron doping the island naturally forms a p--n junction in the
graphene sheet. For example, a doping level of $\sim 3 \times 10^{13}$
electrons per cm$^2$ results in a p--n junction with 800~meV
electrostatic potential barrier.  Unlike in a conventional p--n
junction in graphene, in the case of the junction formed by an
adsorbed organic molecular island we expect that the Klein tunneling
is inhibited, even without an applied external magnetic field.  Here
Klein tunneling is inhibited by the ferromagnetic order that
spontaneously occurs in the molecular island upon doping.  We estimate
that the magnetic barrier in the graphene sheet is around 10~mT.
\end{abstract}



\maketitle

Surface functionalization of graphene using organic molecules is a
promising method to control doping of graphene. Among various organic
molecules, tetrafluoro-tetracyanoquinodimethane (F$_4$TCNQ) is one of
the most intensively studied organic dopants on
graphene \cite{mali,macl,hong,chen,song,wang,macc}.  Since the
electron affinity of F$_4$TCNQ is 5.24~eV\cite{coletti} and the work
function of graphene is 4.6~eV\cite{yu}, the lowest unoccupied
molecular orbital (LUMO) of the molecule lies well below the Dirac
point of graphene and F$_4$TCNQ becomes an efficient p-type dopant
when it is brought into contact with graphene.  Most previous
theoretical studies\cite{pinto,sun,garnica,stradi} of the
F$_4$TCNQ/graphene system have been done considering a single molecule
on a graphene sheet. However, recent experimental study found a
self-assembled F$_4$TCNQ island\cite{tsai} on a graphene substrate
supported by insulating hexagonal boron nitride.  With the formation
of the island, the doping level can be controlled within a much wider
range than for isolated molecules.  For example, depositing naturally
p-type molecular island onto an n-doped graphene would form a p--n
junction in a graphene sheet; so that graphene sheet not covered in
molecules remains n-type while graphene covered with molecules is
p-type.  This indicates that a 2D sheet of F$_4$TCNQ molecules could
be a potential ingredient for fabricating useful graphene-based
electronic devices.

A major obstacle for graphene-based electronics is the inability to
confine Dirac electrons by electrostatic potentials, because of a
unique characteristic of relativistic massless electrons known as
Klein tunneling\cite{katsnelson}.  One way to inhibit the Klein
tunneling is to open a band gap in graphene and consequently change
the linear dispersive electronic property. However, this method
strongly degrades the charge carrier mobility in the graphene layer
which in turn hinders its application for electronic devices.
Alternatively, inhomogeneous magnetic fields can be used to confine
Dirac electrons without opening a band
gap \cite{martino,shytov,rozhkov,young,lu} and reducing the carrier
mobility.

\begin{figure} 
\epsfig{file=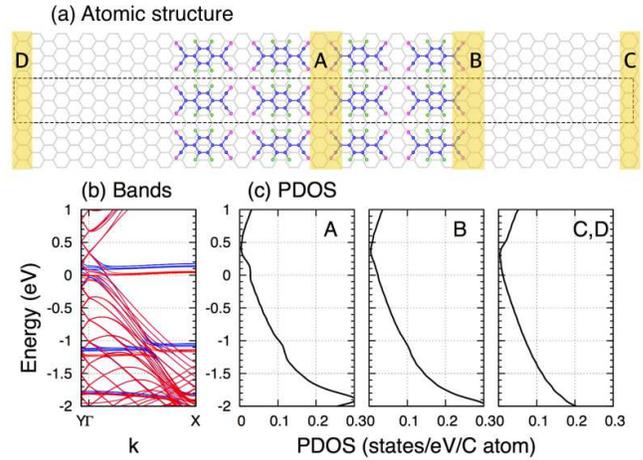,width=8.5cm,clip=}
\caption{(a) Atomic structure of an F$_4$TCNQ ribbon/graphene system.
  Graphene is shown with grey lines. Blue, pink, and green dots
  represent C, N, and F atoms constituting F$_4$TCNQ molecules,
  respectively.  The dotted box shows the supercell used in our study.
  (b) Spin-resolved band structure with the Fermi level at zero energy
  and (c) projected density of states of the undoped case for region
  A, B, C, D shown in yellow.  Majority and minority spins are
  indicated by red and blue colors, respectively.
\label{fig_1}
}
\end{figure}

In this letter, we study the electrical and magnetic properties of a
graphene sheet covered with a molecular F$_4$TCNQ ribbon
[Fig.~\ref{fig_1}(a)]. We find that with electron doping the graphene
sheet indeed forms a p--n junction.  At the same time we find that the
Klein tunneling through this p--n junction is reduced by the
ferromagnetic moment formed on the F$_4$TCNQ molecules.

Our first-principles calculations are based on the use of {\it
ab-initio} norm-conserving pseudopotentials\cite{hamann} and the
Perdew-Burke-Ernzerhof-type\cite{perdew} generalized gradient
approximation to the density functional theory as implemented in the
SIESTA code\cite{sanchez}.  The electronic wavefunctions are expanded
using pseudoatomic orbitals (double-$\zeta$ polarization).  The charge
density cutoff energy is 600~Ry and 32$\times$2$\times$1 $k$-point
sampling is used.  To rule out the undesired interaction arising from
the periodically arranged layers along the out-of-plane $z$ direction,
we placed $\sim$100 \AA~vacuum gap between layers and included a
dipole correction in all calculations.

Our periodic computational unit cell is shown with a dotted line in
Fig.~\ref{fig_1}(a) and it contains a large graphene sheet with 288
carbon atoms.  In order to simulate the finite size effect of the
molecular island we cover half of the graphene sheet with densely
packed F$_4$TCNQ molecules in the shape of a four-molecule wide ribbon
with a width of 47~\AA.  The molecular ribbon (island) is periodically
repeated along the perpendicular direction.  The molecule is oriented
in the ribbon so that its shorter axis is along the periodically
repeated direction of the ribbon.  Each F$_4$TCNQ molecule contains 20
atoms, so the total number of atoms in the computational unit cell is
368.

We first focus on the charge-neutral (undoped) case where the total
ionic and electronic charge in the unit cell is zero.  However, our
calculation still allows for a charge transfer from the graphene sheet
to the molecular island, as long as total charge remains zero.  The
calculated band structure for the undoped case is shown in
Fig.~\ref{fig_1}(b).  The nearly flat bands near the Fermi level and
$\sim$1~eV below the Fermi level originate from the molecular states,
while the dispersive bands originate from the graphene sheet.  We find
that the graphene bands are spin-degenerate and the Dirac point is
located 350~meV above the Fermi level since these organic molecules
act as p-donors. To see the change of electronic properties in the
graphene sheet depending on whether or not a ribbon is on it, we
calculate the projected density of states (PDOS) for three areas:
underneath the ribbon, the edge of the ribbon, and the area without a
covered ribbon [Fig.~\ref{fig_1}(c)]. We find nearly no variation of
the Dirac point in the three areas.  Therefore all parts of graphene
sheet in our periodic super-cell calculation are nearly equally
p-doped by the molecular ribbon.  We expect that with a wider
super-cell, eventually graphene sheet would become neutral far away
from the molecular ribbon.

The ground state of the neutral system is very weakly ferromagnetic.
The energy difference between the ferromagnetic and nonmagnetic states
is only 3~meV per one F$_4$TCNQ molecule. The flat band originating
from the LUMO states of the F$_4$TCNQ molecule is split into one with
the majority spin at the Fermi level and the other with the minority
spin at 100~meV higher than the Fermi level [see Fig.~\ref{fig_1}(b)].

\begin{figure} 
\epsfig{file=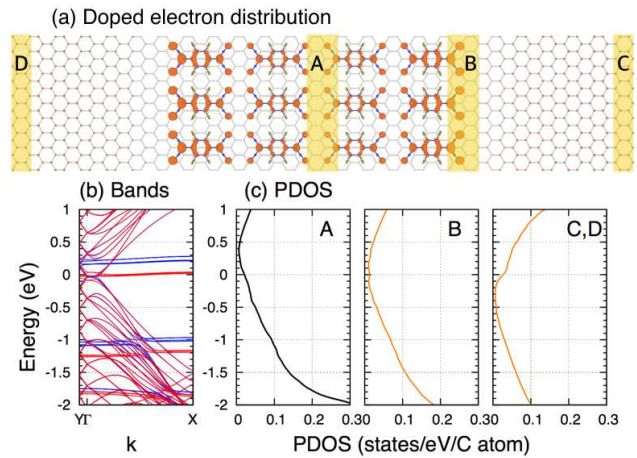,width=8.5cm,clip=}
\caption{(a) The spatial distribution of doped electrons when the
  system is doped with two electrons per cell. The isosurface with
  electron density of 2.7 nm$^{-3}$ is shown with an orange surface.
  (b) Spin-resolved band structure and (c) projected density of states
  of the electron-doped case for region A, B, C, D are shown in
  yellow.  Majority and minority spins are indicated by red and blue
  colors, respectively.
\label{fig_2}
}
\end{figure}

Now we focus on the electron-doped case. To explore the effect of
doping, we added 2 extra electrons in our computational unit cell that
contains 4 molecules.  We estimated the average electron density to be
$2.7 \times 10^{13}$ electrons/cm$^2$ by dividing the added charge by
the area of the computational unit cell.  After relaxation of the
added charge we find that graphene underneath the island remains
unaffected.  Namely, graphene under the molecular island remains
p-doped even with excess electrons in the unit cell [see
  Fig.~\ref{fig_2}(a)].  Instead, excess electron charge accumulates
in the molecular island and in the graphene that is not covered with
molecules.  We find that about 40~\% of the inserted electron charge
accumulate on the ribbon while the remaining 60~\% of the charge goes
into the uncovered part of the graphene.

The excess charge in the uncovered graphene causes its Dirac point to
sink below the Fermi level and it therefore becomes n-type doped.
Since the graphene underneath the ribbon remained p-type we conclude
that the electron doping of the F$_4$TCNQ island on graphene forms a
p--n junction in the graphene sheet.  This charge configuration is
consistent with the fact that the positively charged molecular island
repels electrons from the graphene underneath the island.

The projected density of states analysis reveals that in the uncovered
graphene the energy difference between the Dirac point and the Fermi
level shifts by 800~meV [Fig.~\ref{fig_2}(c)] relative to the neutral
case.

Although this molecular island forms an effective electrostatic
potential barrier, this is not sufficient to confine the Dirac
electrons because of the Klein tunneling effect mentioned earlier.
Nevertheless, the Klein tunneling is inhibited in the F$_4$TCNQ
ribbon/graphene system because electron doping changes not only the
electronic properties but also the magnetic properties of the system.

\begin{figure*} 
\epsfig{file=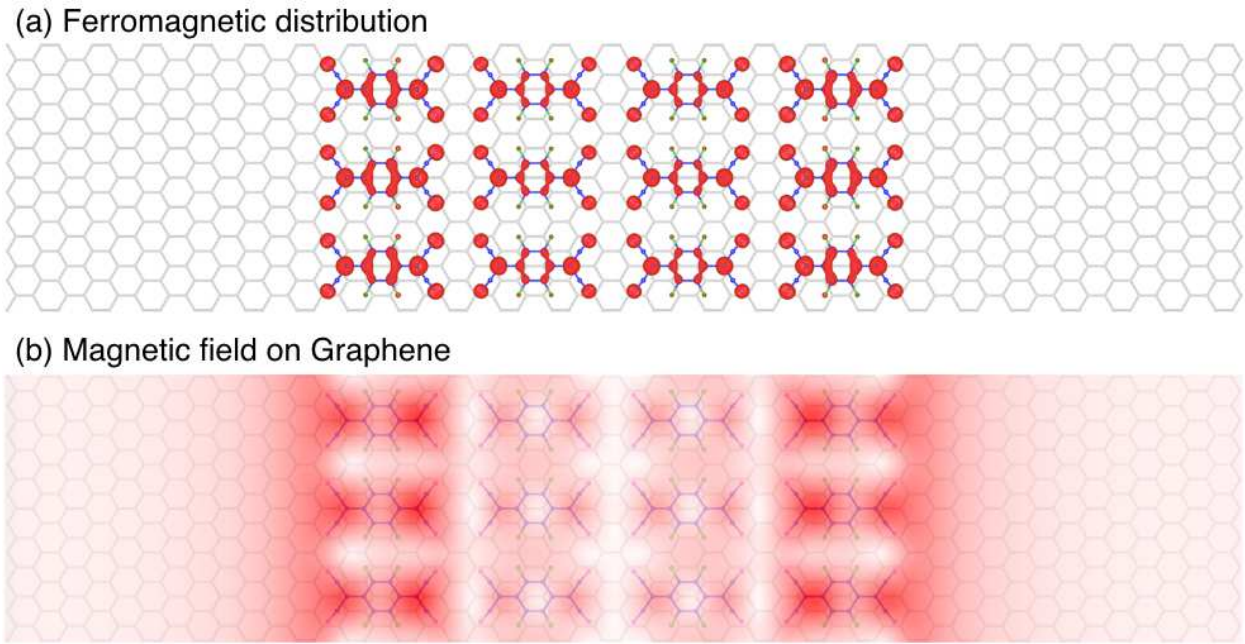,width=16.6cm,clip=}
\caption{(a) Charge density difference between the majority spin and
  minority spin configurations. The isosurface with electron density
  of 6.7 nm$^{-3}$ is shown in red.  (b) The magnitude of the magnetic
  field on graphene. Both figures are for the case when the system is
  doped with two electrons per cell.
\label{fig_3}
}
\end{figure*}

In addition to the formation of the p--n junction, electron doping
enhances the stability of ferromagnetic state and increases the
ferromagnetic moment in F$_4$TCNQ ribbon.  When 2 electrons are added
in the computational unit cell (corresponding to $2.7 \times 10^{13}$
electrons/cm$^2$), the energy difference between the ferromagnetic and
nonmagnetic states is increased from 3~meV/molecule to
21~meV/molecule. The net ferromagnetic moment is calculated to be
0.85~$\mu_B$/4 molecules for the undoped case and 1.81~$\mu_B$/4
molecules for the 2-electron-doped case.

The physical origin of the enhanced magnetism upon electron doping for
this system can be attributed to the Coulomb interaction within the
molecule.  Since the states of the F$_4$TCNQ molecules are spatially
localized within each molecule [Fig.~\ref{fig_3}(a)], the electrons in
the ribbon can be described using a Hubbard model. In the Hubbard
model, the energy gain due to spin polarization is proportional to the
square of the total number of electrons.  According to the theory,
when the total number of electrons in the ribbon increases from 0.85
to 1.81, the energy gain with full polarization is enhanced a factor
of 4.5, which is comparable to the calculated result.

To make a quantitative prediction of how the Klein tunneling can be
inhibited by the magnetism of F$_4$TCNQ ribbon, we calculate the
magnetic field on a graphene sheet when the net ferromagnetic moment
in the super-cell is 1.81 $\mu_B$.  For the calculation, the following
equation is used,
\begin{equation}
\label{}
{\bf B}({\bf r})=\nabla \times \frac{\mu_{0}}{4\pi} \iiint_{all} \frac{{\bf m}({\bf r'}) \times ({\bf r}-{\bf r'})}{\left | ({\bf r}-{\bf r'})^3 \right |} dV'
\end{equation}
where $\left | \iiint_{cell} {\bf m}({\bf r'}) dV' \right | = 1.81
{\mu}_{B}$ and ${\bf m}({\bf r})$ is proportional to the difference
between the majority and minority spin-polarized charge density
computed from first-principles.  The resulting calculated magnetic
field $|{\bf B} ({\bf r})|$ in the graphene sheet is drawn using a
color plot in Fig.~\ref{fig_3}(b).  The maximum value ($\sim$5 mT)
appears under the edge of the ribbon.  The field below the inner
molecules is reduced to 2.2~mT and it becomes less than 0.5~mT away
from the ribbon.  We expect that with a wider super-cell, the magnetic
field away from the ribbon would eventually go to zero.  The
difference in the magnetic field between the edge and the interior of
the ribbon is caused by the variation of the magnetic moment along the
ribbon.  For example, magnetic moment on the molecule at the ribbon
edge is 1.4 times larger than in the interior.

If the system is doped with even more electrons (approximately $5.3
\times 10^{13}$ electrons/cm$^2$), the net ferromagnetic moment can be
increased up to 4~$\mu_B$. With this maximum moment, Dirac electrons
passing under the F$_4$TCNQ ribbon will encounter a $\sim$10~mT
magnetic field barrier.  If the height of the magnetic barrier is
assumed to be constant within the ribbon, the magnetic length $l_B =
\sqrt{\hbar/e B}$ becomes 250~nm.  When the energy of incoming state,
$\varepsilon$, satisfies the following condition
\begin{equation}
\varepsilon \frac{l_B}{\hbar v_f} \leq \frac{d}{l_B}
\end{equation}
where $v_f$ is the Fermi velocity ($\simeq 1 \times 10^6$~m/s) and
$2d$ is the width of a ribbon, the incoming state is reflected
regardless of the incidence angle \cite{martino}.  So, if the width of
the F$_4$TCNQ ribbon is extended to 500~nm, then every incoming state
with energy less than 2.5~meV is totally reflected at the boundary
with p--n junction configuration.

Moreover, an unique inhomogeneous magnetic and electrostatic barrier
geometry in one dimension could be realized by applying homogeneous
perpendicular weak magnetic field on the entire system. The directions
of the magnetic field inside and outside F$_4$TCNQ ribbon can either
be parallel or anti-parallel. From previous theoretical
studies\cite{rozhkov,park,ghosh,oroszlany}, robust one-dimensional
conducting edge states are predicted for such a barrier geometry.
Those edge states have a possibility to be realized in F$_4$TCNQ
ribbon/graphene system, and this system may show a very large
magnetoresistance behavior.

In conclusion, we studied the electric and magnetic properties of an
F$_4$TCNQ ribbon/graphene system. We find that, when the system is
doped with extra electrons, only the uncovered part of the graphene is
doped.  This lowers the Dirac point energy below the Fermi level
making a p--n junction configuration in the graphene sheet.
Furthermore, we find that electron doping induces tunable
ferromagnetism in the ribbon. When extra electrons flow into the
system, the ferromagnetic moment in the ribbon is increased and that
moment produces the magnetic barrier high enough to confine Dirac
electrons. Our findings reveal the possibility of tunable
electrostatic and magnetic barriers in graphene, which could be
effective for inhibiting Klein tunneling in graphene-based electronic
devices.

This work was supported by National Science Foundation Grant
No. DMR15-1508412 (electronic structure calculation) and by the
Director, Office of Science, Office of Basic Energy Sciences,
Materials Sciences and Engineering Division, U.S. Department of Energy
under Contract No. DE-AC02-05CH11231, within the SP2 Program (magnetic
structure calculation).  Computational resources have been provided by
the DOE at Lawrence Berkeley National Laboratory's NERSC facility.

\end{document}